\documentclass[article]{rstransa}
\usepackage{graphicx}
\usepackage{caption}
\usepackage{comment}
\usepackage{authblk}
\usepackage{amsmath}
\usepackage{amssymb}
\usepackage[normalem]{ulem}
\usepackage{tikz}
\DeclareRobustCommand\full  {\tikz[baseline=-0.6ex]\draw[thick] (0,0)--(0.5,0);}
\DeclareRobustCommand\dotted{\tikz[baseline=-0.6ex]\draw[thick,dotted] (0,0)--(0.54,0);}
\DeclareRobustCommand\dashed{\tikz[baseline=-0.6ex]\draw[dashed] (0,0)--(0.54,0);}

\DeclareRobustCommand\chain {\tikz[baseline=-0.6ex]\draw[thick,dash dot] (0,0)--(0.54,0);}

\begin{document}

\title{Taylor-Couette flow of hard-sphere suspensions: Overview of current understanding}

\author{
Lina Baroudi$^{1}$, Madhu V. Majji$^{2}$, Stephen Peluso$^{1}$, and Jeffrey F. Morris$^{3}$}

\address{$^{1}$Department of Mechanical Engineering, Manhattan College, Bronx, NY 10471, USA \\$^{2}$Department of Chemical Engineering, Massachusetts Institute of Technology, Cambridge, MA 02139, USA $^{3}$Levich Institute and Department of Chemical Engineering, CUNY City College of New York; New York, NY 10031, USA}


\corres{Lina Baroudi\\
\email{lina.baroudi@manhattan.edu}}

\begin{abstract}
\sloppy

Although inertial particle-laden flows occur in a wide range of industrial and natural processes, there is both a lack of fundamental understanding of these flows and continuum-level governing equations needed to predict transport and particle distribution. Towards this effort, the Taylor-Couette flow (TCF) system has been used recently to study the flow behavior of particle-laden fluids under inertia. This article provides an overview of experimental, theoretical, and computational work related to the TCF of neutrally buoyant non-Brownian suspensions, with an emphasis on the effect of finite-sized particles on the series of flow transitions and flow structures. Particles, depending on their size and concentration, cause several significant deviations from Newtonian fluid behavior, including shifting the Reynolds number corresponding to transitions in flow structure and changing the possible structures present in the flow. Furthermore, particles may also migrate depending on the flow structure, leading to hysteretic effects that further complicate the flow behavior. The current state of theoretical and computational modeling efforts to describe the experimental observations is discussed, and suggestions for potential future directions to improve the fundamental understanding of inertial particle-laden flows are provided.  

\fussy

\end{abstract}
\maketitle

\section{Introduction}
Particle-laden flows are widely encountered in industrial applications, such as cement mixing, slurry transport, and pharmaceutical processes, as well as in natural phenomena like mud flow and sediment transport. Different flow regimes can be observed in these contexts, ranging from the Stokes regime to flows with various inertia levels, including the turbulent regime. Thus, predicting the behavior of suspensions under varied flow conditions requires consideration of inertial flow phenomena. However, assessing the effect of particle-scale inertia on rheology is challenging as inertial effects grow with length scale; finite microscale inertia may be associated with large macroscopic inertia, which introduces instability and may lead to secondary flow and turbulence. Recent studies on inertial instabilities of particle-laden Taylor-Couette flow have shown that particles play a fundamental role, and the observed dynamical behavior cannot yet be explained by effective medium approaches~\cite{majji2018inertial_JFM,majji2018inertial,ramesh2019suspension,gillissen2019taylor,ramesh2020interpenetrating,dash2020particle,baroudi2020effect,kang2021flow,moazzen2022torque,singh2022counter}. 

\begin{figure}
\centering
\includegraphics[scale=0.90]{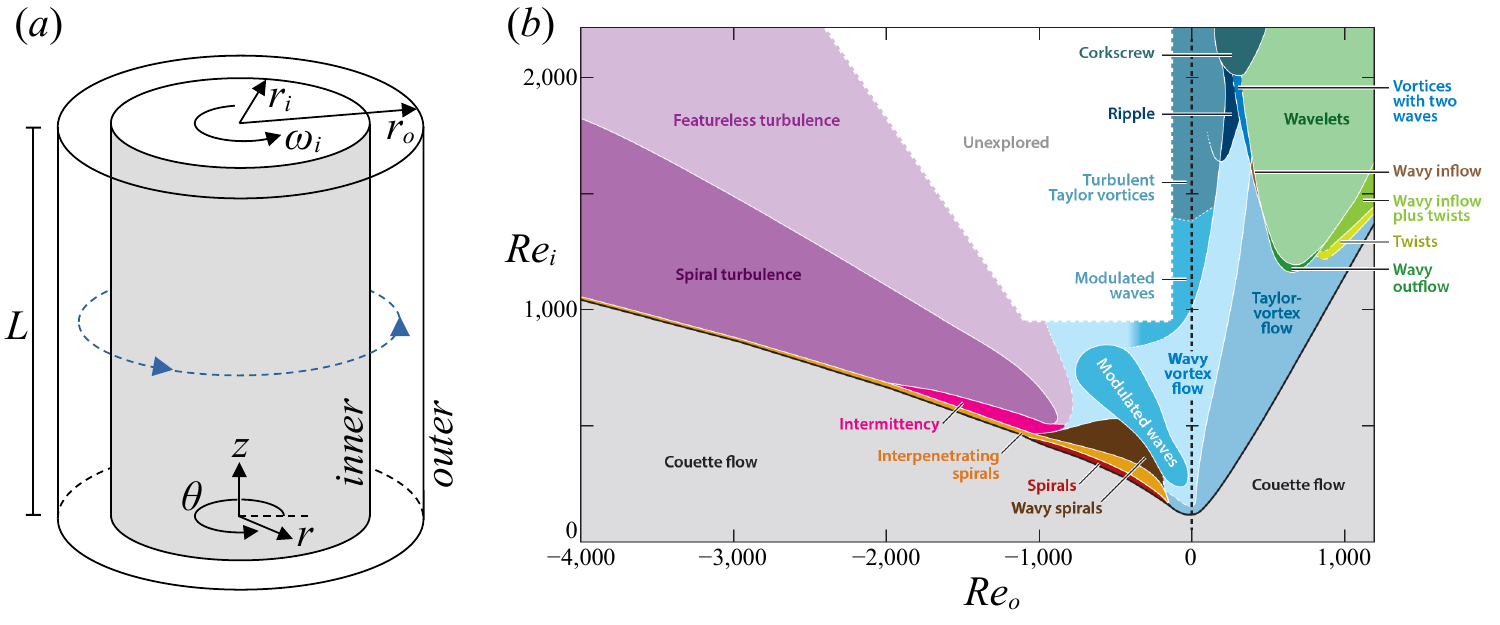}
\caption{(a) Sketch of Taylor-Couette flow (b) Flow structures of pure fluid in an annulus between two independently rotating concentric cylinders with radius ratio $\eta = 0.88$. Adapted from Andereck {\em et al.}~\cite{andereck1986flow}. Colored version reprinted from~\cite{grossmann2016high}.}\label{fig:setup-map}
\end{figure}

The Taylor-Couette flow (TCF), i.e., the flow between differentially rotating concentric cylinders (Fig.~\ref{fig:setup-map}(a)), has played a significant role in developing  fundamental understanding of fluid dynamics. It is considered a paradigm for hydrodynamic stability and transition studies. Since the original experiments of Mallock~\cite{mallock1889iv} and Couette~\cite{couette1890etudes} and the seminal analysis and experiments of Taylor~\cite{taylor1923viii}, the flow transitions of a Newtonian fluid in the Taylor-Couette geometry have been extensively studied in various experimental, theoretical, and numerical works~\cite{taylor1923viii,chandrasekhar1958stability,coles1965transition,mullin1980transition,lorenzen1983end,andereck1986flow,wereley1998spatio,czarny2002spiral,hoffmann2004spiral,grossmann2016high}. At low rotation rates of the cylinders the realized flow is the circular Couette flow (CCF) which is unidirectional in the azimuthal direction and only has radial dependence. As the rotation rates of the cylinders increases, the CCF transitions to various flow structures depending on the rotation rates of the inner and outer cylinders~\cite{taylor1923viii,coles1965transition,andereck1986flow} and the rheological properties of the fluid~\cite{muller1989purely,larson1990purely}. Taylor~\cite{taylor1923viii}, in 1923, was the first to carry out calculations of the linear instability CCF to the onset of axisymmetric vortices. His experimental measurements were in remarkable agreement with his calculations, one of the major success stories of linear stability theory. In later work, Andereck \textit{et al.}~\cite{andereck1986flow} experimentally mapped a much more complete spectrum of flow transitions and flow structures corresponding to the Reynolds numbers based on inner and outer cylinder rotation rates: $Re_i=\rho \delta\omega_i r_i/\mu$  and $Re_o=\rho \delta\omega_o r_o/\mu$, where $\delta$ is the annular gap, $r_i$ and $r_o$ are the radii of the inner and outer cylinders, $\omega_i$ and $\omega_o$ are the rotation rates of inner and outer cylinders and $\rho$ and $\mu$ are the density and viscosity of the Newtonian fluid. Various spatial and temporal flow structures, including Taylor vortices, spirals, modulated waves, and many other intriguing flow states, are depicted in the phase diagram of Andereck reproduced in Fig.~\ref{fig:setup-map}(b). The rich dynamical behavior of TCF, its simple geometry, and the applicational relevance of understanding particle motion in this flow~\cite{kroner1988dynamic, moore1995axial, wereley1999inertial, dherbecourt2016experimental,rida2019experimental} make it an excellent tool for understanding the role of inertia and flow instabilities in particle-laden fluids. Additionally, linear stability theory provides valuable guidance to the flow transitions in TCF~\cite{chandrasekhar1958stability}. The ability to predict the effect of particles on the alteration of existing instabilities known for pure fluids and onset of turbulence would  significantly impact the efficiency of processes that involve mixing and transport of suspensions.

 The simplest type of suspension, which consists of mono-disperse rigid neutrally-buoyant spheres in a Newtonian fluid, displays intricate rheological properties. The rigidity of the particles prevents deformation under flow and causes the generation of flow disturbances. The interaction of these disturbances with one another, with the boundaries of the flow geometry, and with the background shear gradients causes macroscopic changes in viscosity and normal stresses, and modifications of flow profiles and their stability characteristics. For a dilute suspension ($\phi \ll1$) of solid spheres, the effective viscosity of suspension ($\mu_s$) can be determined by Einstein's~\cite{Einstein1906} relation, $\mu_s(\phi)/\mu = 1+(5/2)\phi$, where $\phi$ is the particle volume fraction. Taylor extended Einstein's work to liquids containing small drops of another liquid in suspension~\cite{taylor1932viscosity}; a few decades later, Batchelor \& Green~\cite{batchelor1972hydrodynamic} considered the first effects of particle interactions and extended Einstein's relation to the second order in $\phi$, but still limited to $\phi\ll1$. For larger $\phi$, Batchelor~\cite{batchelor1970stress} proposed a constitutive equation for the bulk stress in the suspension ($\Sigma$) which consists of stress contribution from the fluid phase ($\mathbf{\Sigma_{f}}$) and the particles ($\mathbf{\Sigma_{p}}$): $\mathbf{\Sigma} = \mathbf{\Sigma_{f}}+\mathbf{\Sigma_{p}}$, with $\mathbf{\Sigma_{f}} = \frac{1}{V}\int_{v_{f}}[-p\mathbf{I}+\mu(\mathbf{\nabla u}+\mathbf{\nabla u^{T})}]dv$ and $\mathbf{\Sigma_{p}} = \frac{1}{V}\int_{v_{p}}\mathbf{\sigma} dv +\frac{1}{V} \int_{V} \rho \mathbf{u}^\prime\mathbf{u^\prime} dv$. Here $p$ is the pressure, $\mathbf{u}$ is the velocity, $v_f$ and $v_p$ are the parts of fluid phase and the solid phase of the control volume $V$ over which the stress components are averaged to arrive at the macroscopic continuum description of the suspension flow. The particle stress ($\mathbf{\Sigma_{p}}$) consists of contributions from the local stress tensor $\mathbf\sigma$ in the solid phase and the \textit{Reynolds stress} due to velocity fluctuations (${\mathbf{u^\prime}}$) in the control volume. The velocity fluctuations are generated by the individual disturbances in the  flow caused by the particles and their interactions. Given their seminal contributions to this flow and to suspension dynamics, had Taylor or his student, Batchelor, addressed the influence of suspended particles on inertial flow transitions in TCF, one may wonder whether we would have any work left to do on this problem!

The discussion in this review will focus on inertial instabilities of neutrally buoyant suspensions of non-Brownian spheres with hard interactions in Newtonian liquid flowing between concentric cylinders. The non-Newtonian behavior will thus arise from the presence of the particles. Flow transitions in TCF of particle suspension in a more complex fluid are discussed elsewhere~\cite{lacassagne_2021}. The suspension consists of hard spheres of diameter $d_p$ suspended in a fluid of viscosity $\mu$, the density of both fluid and particles for a neutrally buoyant suspension is $\rho$. Given two concentric cylinders of inner radius $r_i$, outer radius $r_o$, annular width $\delta = r_o-r_i$, and inner cylinder height $L$, the non-dimensional parameters characterizing the TC apparatus and the materials are the radius ratio $\eta ={r_i}/{r_o}$, the aspect ratio $\Gamma = L/\delta$, the particle volume fraction $\phi$, and the ratio of the annular gap $\delta$ to the particle diameter $\alpha = \delta/d_p$. The particle Reynolds number $Re_p = {\rho \dot{\gamma}d_{p}^{2}}/{4\mu}$ represents the balance of inertial and viscous effects at the particle scale, where $\dot{\gamma}$ is the shear rate. The P\'{e}clet number $Pe = 6\pi\mu_{0}\dot{\gamma}a^{3}/kT$ defines the ratio of shear to thermal motion with thermal energy $kT$, where $k$ is the Boltzmann constant and $T$ is the absolute temperature. For all studies considered in this review, $Re_p$ is relatively small but finite and particles are non-Brownian, so that $Pe\gg1$. The bulk (suspension) Reynolds numbers are defined based on the inner and outer cylinders' angular velocities $Re(\phi) = {\rho \omega_i r_i  \delta}/{\mu_{s}(\phi)}, Re_{o}(\phi) = {\rho \omega_o r_o  \delta}/{\mu_{s}(\phi)}$, respectively, where $\mu_{s}(\phi)$ is the effective viscosity of the suspension as a function of the solid volume fraction $\phi$. The bulk and the particle Reynolds numbers are related as follows $Re_{p}\sim (d_p/2\delta)^{2}Re$.

\begin{figure}
  \centering
  \includegraphics[scale=1.0]{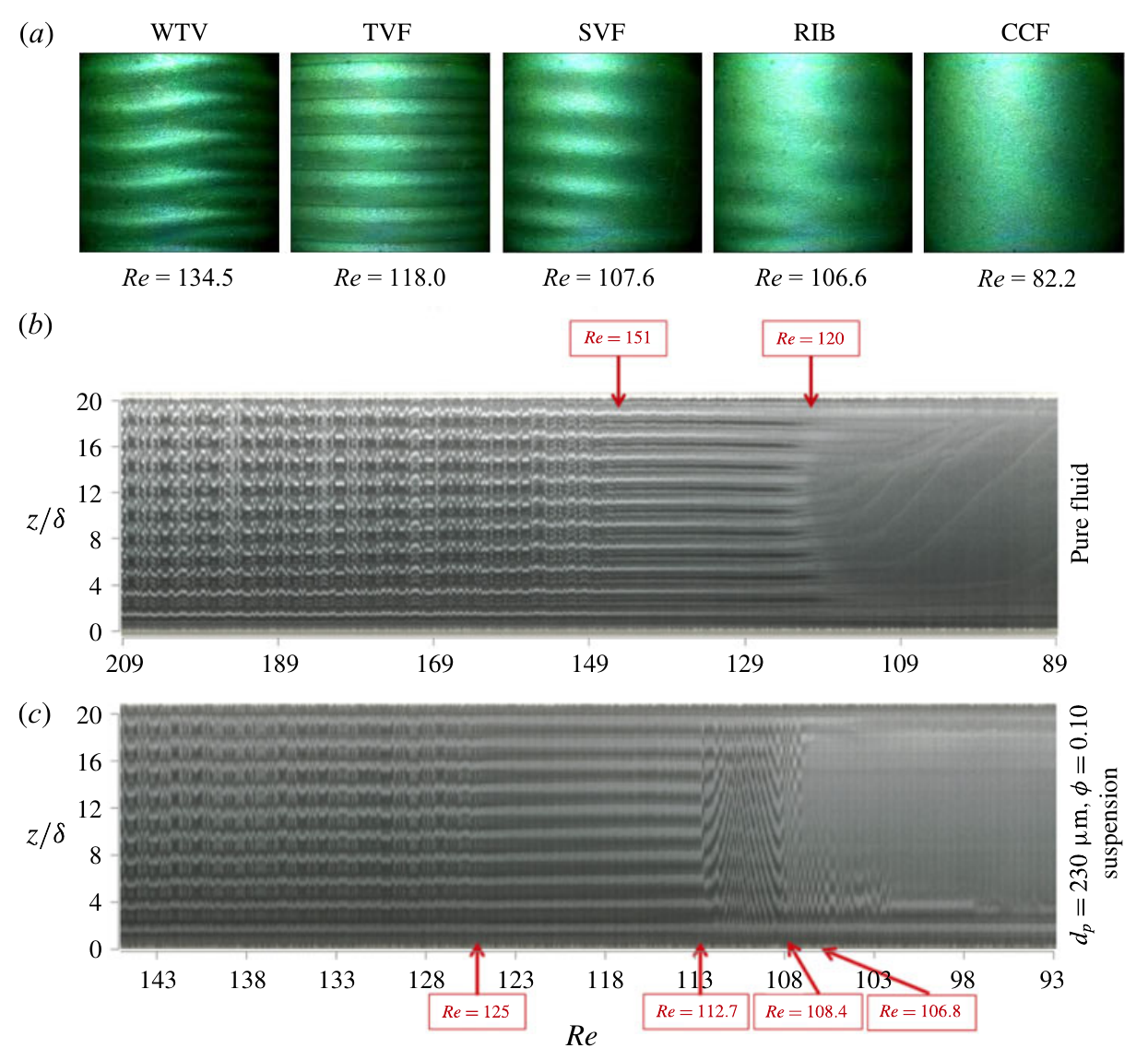}
  \caption{(a) Photographs of flow structures for a $\phi = 0.10$ suspension with a gap-to-particle size ratio of $\alpha = \delta/d_p = 30$ observed during a decreasing-$Re$ ramp, from left to right: wavy Taylor vortices (WTV), Taylor vortex flow (TVF), spiral vortex flow (SVF), ribbons (RIB), and circular Couette flow (CCF) at $Re = 134.5, 118, 107.6, 106.6,$ and $82.8$, respectively. Space-time diagrams for (b) pure fluid, and (c) $\phi = 0.1$ suspension of $\alpha = 30$ from decreasing-$Re$ experiments. Reprinted from Majji~\textit{et al.}~\cite{majji2018inertial_JFM}. }\label{fig:Sus_flow_states}
\end{figure}

\begin{figure}
  \centering
  \includegraphics[scale=1.0]{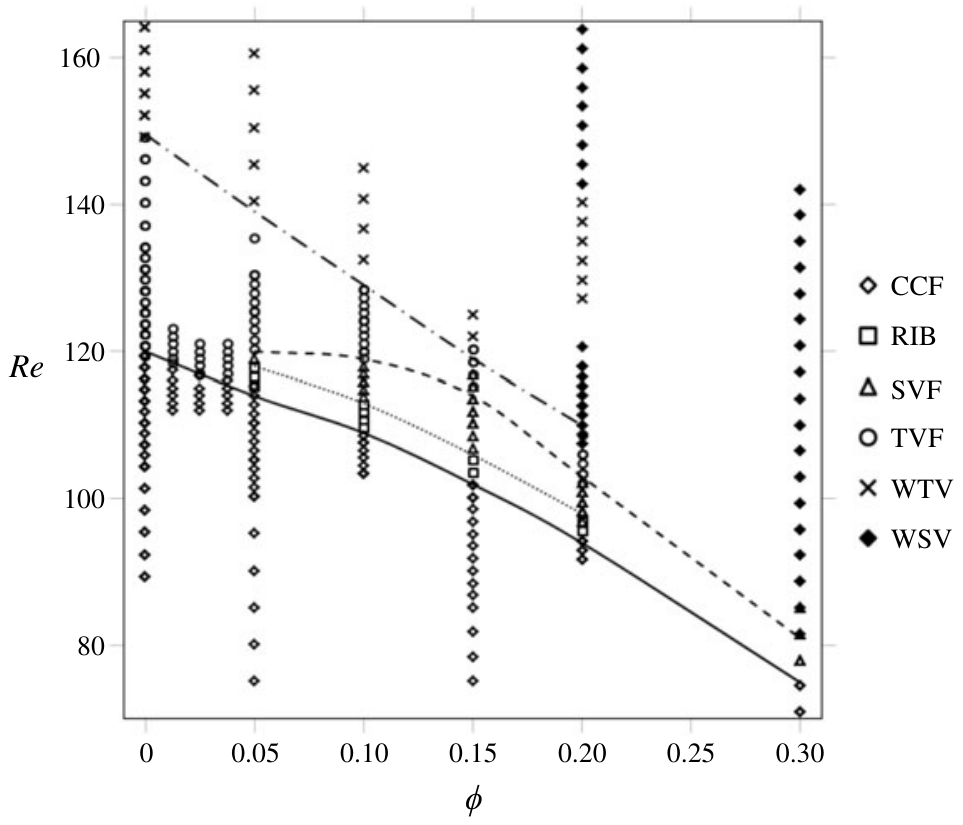}
  \caption{Suspension flow state diagram from decreasing-$Re$ ramp experiments for $0\leq\phi\leq 0.30$ and gap-to-particle size ratio of $\alpha = 30$ in TC apparatus with radius ratio $\eta = 0.877$ and aspect ratio $\Gamma = 20.5$. Lines depicting different flow transitions: \full, CCF-RIB;~\dotted, RIB-SVF;~\dashed, SVF-TVF;~\chain, TVF-WTV. Abbreviations: CCF, circular Couette flow; RIB, ribbons; SVF, spiral vortex flow; TVF, Taylor vortex flow; WTV, wavy Taylor vortices; WSV, wavy spiral vortices. Reprinted from Majji {\em et al.}~\cite{majji2018inertial_JFM}. }\label{fig:suspension_flow_map}
\end{figure}

The remainder of this review covers experimental, modelling, and linear stability analysis studies of particle-laden TCF. Early experimental studies examining lower-order transitions with only inner cylinder rotation are discussed first (\ref{Experiments}\ref{lowOrderRegime}) and used to provide a framework for understanding the differences between Newtonian and suspension flows. Observations from these studies showed that particle distribution plays an important role in these differences and more detailed studies of particle migration follow in section \ref{Experiments}\ref{particleMigration}.  Section \ref{Experiments}\ref{torqueMeasurements} covers inner cylinder torque measurements, as this is a common diagnostic for determining flow state transition behavior. Additionally, torque measurements offer insight to the relationship between the driving torque and the Reynolds number. The complexity of the flow state map is known to increase with counter-rotation and the limited study of suspensions under these conditions is described in section \ref{Experiments}\ref{lowCounterRegime}. The discussion of experimental studies is concluded with recent work on  higher-order turbulent transitions in section \ref{Experiments}\ref{highOrderRegime}.  

Studies employing linear stability analysis and modeling of suspension TCF follow in section \ref{LSA-Modelling} and the key differences in observed behavior between experiments and modelling are noted. The review finishes with an overall perspective on the state of understanding of suspension TCF with suggestions for future work.

\section{Experimental studies}\label{Experiments}

The devices used in these studies fall into two categories:  larger-scale, custom-built devices driven by a stepper or servo motor and smaller-scale, rheometer-based devices.  Table 2 in Moazzen {\em et al.}\textbf{~\cite{moazzen2022torque}} contains a detailed summary of the experimental devices and operating conditions for most of the studies summarized here. 
All experimental studies examined neutrally buoyant, non-Brownian hard sphere suspensions.  
The suspensions were sheared between concentric inner and outer cylinders and the set of flow transitions and flow structures observed were compared against the behavior of a pure Newtonian fluid under similar conditions. 

\subsection{Lower-order transitions driven by inner cylinder rotation} \label{lowOrderRegime}

In the first experimental study of non-dilute particle suspensions in Taylor-Couette flow, Majji, Banerjee, and Morris~\cite{majji2018inertial_JFM} explored the effect of particles on fluid behavior over a concentration range of $0.0 \le \phi \le 0.3$. The inner cylinder rotation rate was nondimensionalized as the suspension Reynolds number, $Re$, and the flow behavior was recorded over a range of $Re$ and $\phi$.  All Newtonian fluids are expected to display identical fluid behavior at a given $Re$; therefore, any deviation from the base-line Newtonian behavior for suspensions was attributed to the particles. In addition, particles of two mean particle sizes, $d_p\approx$ 230 $\mu$m and 70 $\mu$m (yielding $\alpha$ = 30 and 100), were used.

The flow structures and flow transitions were recorded as the degree of inertia was reduced, i.e. down-ramped, in a quasi-steady manner over $165 >Re > 70$. 
From the recorded image sequences, space-time diagrams were constructed by extracting a single line of pixel intensity over the entire length of the annulus and plotting it as a function of time or, equivalently, $Re$.  Fig.~\ref{fig:Sus_flow_states}(a) (reproduced from their work) shows examples of images of different flow state structures, while Fig.~\ref{fig:Sus_flow_states}(b\&c) shows two examples of such space-time diagrams.   

For Newtonian fluids ($\phi$ = 0), as $Re$ was reduced from $Re$ = 165, the flow structure transitioned from wavy Taylor vortices (WTV) into Taylor vortex flow (TVF) at $Re \approx$ 151 and from TVF into CCF at $Re\approx$ 120. Experiments with suspensions reveal three key deviations from this Newtonian behavior due to the presence of particles:

First, several stable, non-axisymmetric flow (NAF) structures appeared, as can be seen in Fig.~\ref{fig:Sus_flow_states}:  ribbons (RIB), spiral vortex flow (SVF), and wavy spiral vortices (WSV). At high concentrations ($\phi$=0.3), the flow transitioned through only non-axisymmetric structures before reaching CCF, bypassing TVF entirely. These non-axisymmetric structures were previously observed only in Newtonian fluids with counter-rotating cylinders~\cite{andereck1986flow}, as seen in Fig.~\ref{fig:setup-map}(b), indicating that the particles alter the qualitative nature of the TCF  dynamical system. In addition, the wave characteristics were similar to those observed in counter-rotating studies:  the wave length in the axial direction was equal to twice the annular gap $\delta$ and wave rotation rate was equal to half of the inner cylinder rotation rate $\omega_i$. From this, the authors concluded that the presence of particles enables the suspension to access NAF structures (at different conditions) from the library of solutions to the Newtonian fluid problem, rather than these being ``new'' flow structures. 

Second, the $Re$ corresponding to each of the flow transitions reduced with an increase in particle concentration, indicating that particles destabilize the flow structures, even for low suspension concentrations that followed the same flow transition sequence as the Newtonian fluid (WTV $\to$ TVF $\to$ CCF). This effect increased with increasing suspension concentration; for example, the $Re$ for transitioning into CCF decreased from $Re\approx 120$ for $\phi =0$ to $Re\approx$75 for $\phi= 0.3$. In addition, the $Re$ range over which the non-axisymmetric flow structures were observed to be stable increased with increase in $\phi$. The observed flow transitions and flow structures for various concentrations at $\alpha =30$ can be seen in the suspension flow state diagram, reproduced from Majji {\em et al.}, in Fig.~\ref{fig:suspension_flow_map}. 

As the increase in viscosity due to the particles is accounted for in the effective suspension viscosity, the change in transition values indicates that the presence of the particles destabilizes the flow and the destabilization effect increases with increase in particle concentration. Given the finite particles size ($\alpha=30$) and inertia in the system ($Re\gg 1$), inertial migration of particles leads to a non-uniform concentration distribution of particles in the annular gap depending on the flow structure, as shown in the experiments of Majji and Morris~\cite{majji2018inertial}, leading to non-uniform bulk viscosity profiles. Linear stability analysis \cite{majji2018inertial_JFM} showed that the non-uniform viscosity profiles that account for inertial migration stabilize the CCF structure, contrary to the experimentally observed destabilization. This indicates that neither taking into account the effective suspension viscosity nor adding the effect of spatially varying viscosity in accordance with the inertially migrated concentration profiles rationalizes this behavior. 

Returning to Majji, Banerjee, and Morris~\cite{majji2018inertial_JFM}, the third main deviation noted between suspensions and Newtonian fluids was hysteresis in transition $Re$, depending on the ramp direction.  A comparison between increasing $Re$ (ramp-up) and decreasing $Re$ (ramp-down) protocols revealed that the $Re$ corresponding to SVF-TVF, TVF-WTV transitions were higher than those of TVF-SVF, WTV-TVF transitions respectively; note that the Newtonian fluid displayed no hysteresis for these transitions. This dependence on ramp direction suggests that particle concentration distribution in the annular region plays an important role in the stability of flow structures. Experiments of Majji and Morris~\cite{majji2018inertial} show that particles in time-invariant structures like CCF and TVF inertially migrate to form specific concentration profiles in the annular region whereas in the time-modulating structures like SVF and WTV particles do not have enough time for inertial migration leading to uniform particle distribution. Hence, in the hysteresis region, the concentration profile impacts on the stability of the flow structure. For example, at any $Re$ in the TVF, WTV hysteresis region, the inertially migrated concentration profile stabilizes the TVF structure and destabilizes the WTV structure.  Particle distribution likely has a significant effect on the differences between Newtonian and suspension flows and studies exploring particle migration in more detail are discussed in subsection \ref{particleMigration}.

The experiments of Majji, Banerjee, and Morris~\cite{majji2018inertial_JFM} showed that the addition of particles to a Newtonian fluid in a TC setup with only inner cylinder rotation leads to the appearance of stable NAF structures, the reduction in flow transition $Re$, and hysteresis in flow transitions. The authors hypothesized that the finite-sized perturbations generated by particles and their non-linear interactions in inertial flows may be driving these non-Newtonian TCF behavior. Since these finite-sized perturbations scale with the particle size, as the particle size gets smaller, the flow behavior should approach that of a Newtonian fluid. Experiments showed that when the particle size was reduced from $d_p\approx 230$ $\mu$m to 70 $\mu$m (from $\alpha= 30$ to 100) for a $\phi = 0.1$ suspension, the size of the non-axisymmetric region was reduced considerably. Moreover, the non-axisymmetric region for the $\alpha=30$ case consisted of SVF and RIB whereas for $\alpha= 100$, it consisted only of a small region of RIB. This strongly supports the hypothesis.


The study of Ramesh, Bharadwaj and Alam~\cite{ramesh2019suspension} used a rheometer-based device and several additional diagnostics to improve characterization of the fluid behavior. In addition to direct visualization of the observable fluid motion at the outer cylinder, an index-matched suspension with tracer particles was used for PIV measurement of fluid velocities and torque on the inner cylinder was measured. Although smaller in scale than Majji {\em et al.}~\cite{majji2018inertial_JFM}, the two studies used similar $Pe$ and $Re_p$ ranges and similar gap-to-particle ratios. Ramesh {\em et al.} observed similar effects of particles on the flow behavior to Majji {\em et al.}.  In particular, the critical $Re$ for onset of the primary instability of CCF was reduced with increasing concentration and this primary transition occurred between TVF and CCF for $\phi \le 0.05$ (ramp-down protocol). For higher $\phi$, the work confirmed that the primary bifurcation was between CCF and a non-axisymmetric state, with SVF the finding in this work. The RIB state described by Majji \textit{et al.} was not observed; the authors suggest that this may be related to the smaller aspect ratio $\Gamma$ in their work. Overall, the behavior of the flow was largely similar between the two devices, suggesting that the dimensionless characterization with these parameters is sufficient.  

 An important new finding in Ramesh {\em et al.} (shown in their Fig. 5) is that for ramp-up protocol, there are regimes where two states persist within the annulus simultaneously, typically separated vertically over the height of the apparatus.  Two examples of these long-lived co-existing states were (i) WTV+TVF, arising from a primary bifurcation from CCF for $0.05<\phi\le 0.13$ in the sequence CCF $\rightarrow$ WTV+TVF $\rightarrow$ TVF $\rightarrow$ WTV and (ii) TVF+SVF, arising as a secondary bifurcation in the sequence CCF $\rightarrow$ SVF $\rightarrow$ TVF+SVF $\rightarrow$ TVF $\rightarrow$ WTV for $0.13 < \phi \le 0.25$.
 
\begin{figure}
  \centering
    \includegraphics[scale=1.0]{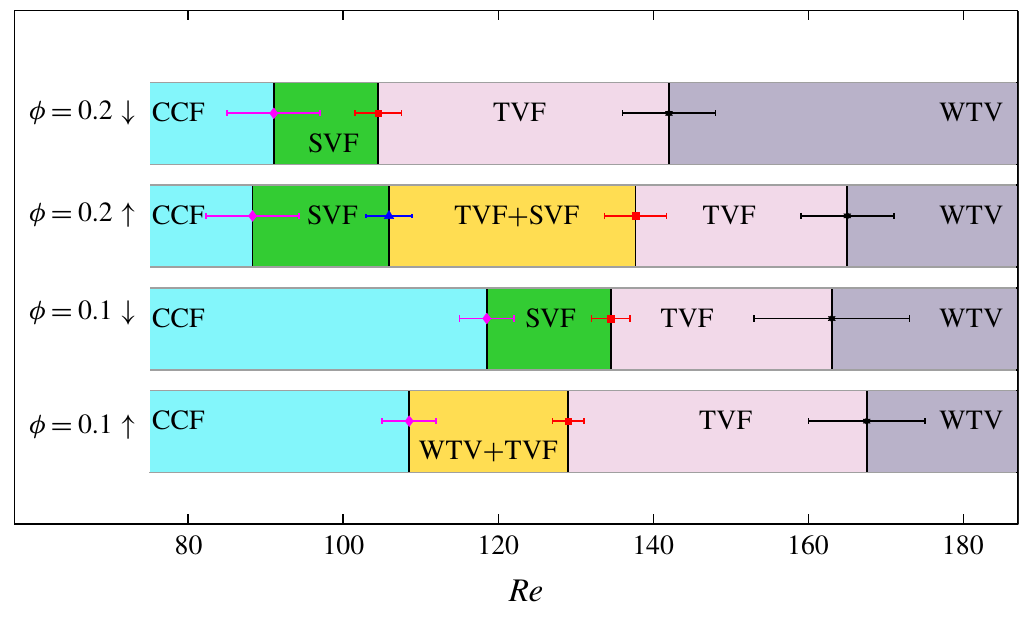}
  \caption{Flow state map in TCF of suspension at $\phi = 0.1$ and $0.2$ with gap-to-particle size ratio $\alpha = 37.5$. The map shows a clear dependence of state boundaries and observed flow states on the $Re$-ramping direction, \textit{i.e.,} hysteresis in the flow transition behavior. Abbreviations: CCF, circular Couette flow; SVF, spiral vortex flow; TVF, Taylor vortex flow; WTV, wavy Taylor vortices. Reprinted from Ramesh {\em et al.} ~\cite{ramesh2019suspension}.}\label{fig:Ramesh}
\end{figure}

 This behavior indicates the hysteresis of transitions in suspension TCF, and Ramesh {\em et al.} develop this issue and its difference relative to a Newtonian fluid, with Fig.~\ref{fig:Ramesh} (reproduced from their work), providing a nice indication of this behavior in a simple-to-follow pattern map for $\phi = 0.1$ and 0.2 as a function of $Re$. The hysteresis under variation of $\Gamma$ and $r_i/r_o$ is also developed.

Ramesh and Alam~\cite{ramesh2020interpenetrating} found new flow states using a similar rheometer-based device with a larger $\Gamma$ compared to their previous work~\cite{ramesh2019suspension}. The major new observation discussed in this study was the appearance of interpenetrating spiral vortices (ISVs) for suspensions with $\phi \ge 0.10$.  In ISV flow, spiral vortices form at the mid-height of the cylinder and propagate both up and down. These vortices appeared in both ramp-up and ramp-down tests between co-existing TVF-SVF and WVF. ISVs had been observed in previous studies with pure fluids in counter-rotating cylinders, but not in suspensions with only inner cylinder rotation; this observation further demonstrates that particle addition increases the states accessible to the flow. 

\subsection{Particle Migration} \label{particleMigration}
One aspect that differs significantly from Newtonian fluids, owing to the finite size of the particles, is that particles can migrate and lead to spatial variation of the effective fluid properties. At finite Reynolds number in dilute suspension flows, the interaction of particle disturbances with the background flow and the solid boundaries leads to inertial migration of the particles to specific locations depending on the carrying flow considered~\cite{segre1962behaviour1,  segre1962behaviour2,ho1974inertial,schonberg1989inertial, asmolov1999inertial,matas2009lateral}. At intermediate and high concentrations, shear-induced migration due to particle interactions~\cite{Morris1999}, along with inertial migration, have been shown in pressure-driven flow to lead to nonuniform distribution of particles and modified velocity profiles~\cite{han1999particle}. 

\begin{figure}
  \centering
    \includegraphics[scale=0.8]{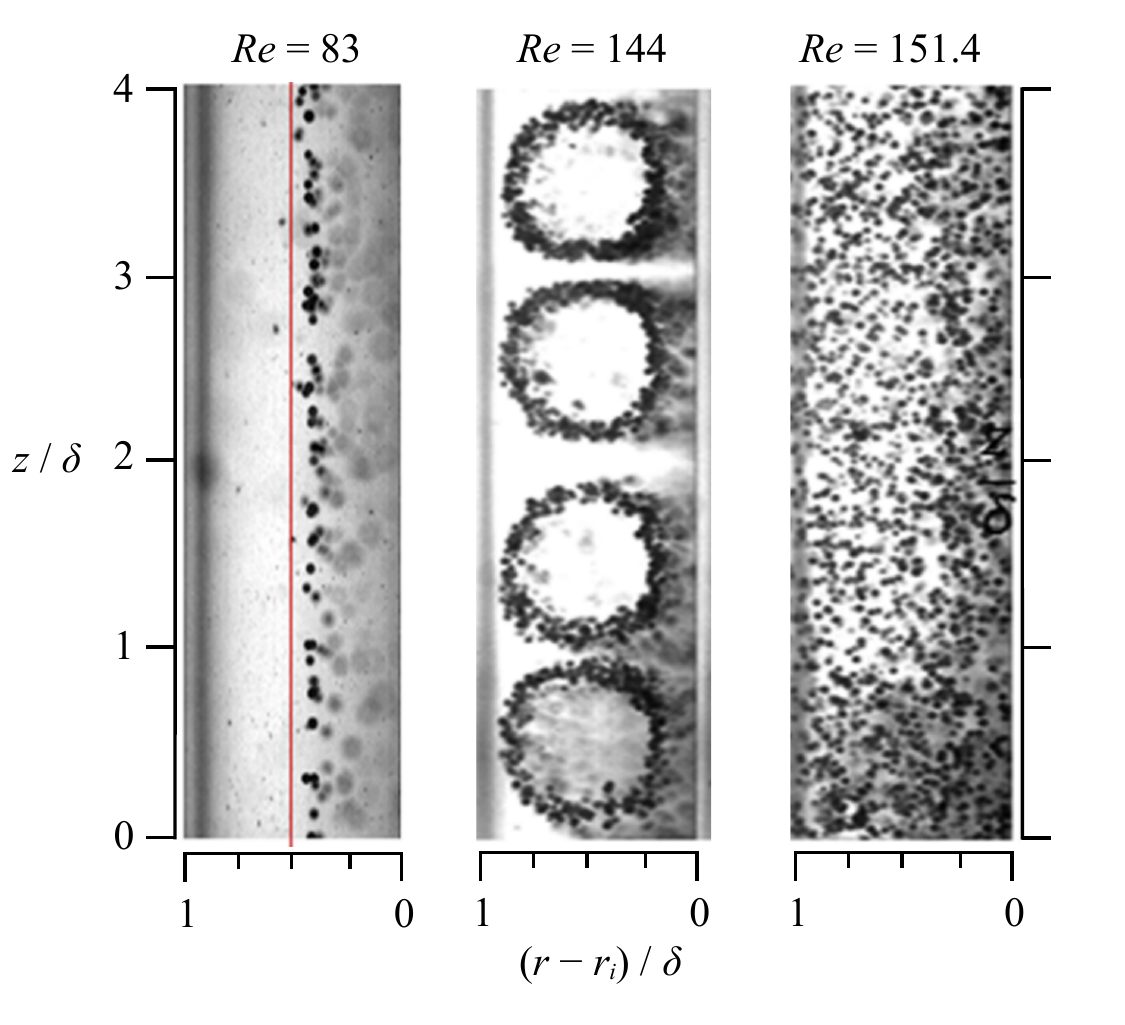}
  \caption{Inertial migration of particles in different flow regimes in an $r-z$ plane for $\phi = 0.001$ and gap-to-particle size ratio of $\alpha = 20$ in TC geometry with radius ratio $\eta = 0.877$, and aspect ratio $\Gamma = 20.5$. Dark and sharp spots are the particles in the plane of interest. Flow structures, from left to right: circular Couette flow (CCF) at $Re = 83$, Taylor vortex flow (TVF) at $Re = 144$, and wavy Taylor vortices (WTV) at $Re = 151.4$.  
  Adapted from Majji and Morris~\cite{majji2018inertial}.}\label{fig:migration-dilute}
\end{figure}

In Taylor-Couette geometry, Halow \textit{et al}.~\cite{halow1970experimental} observed that neutrally-buoyant spheres in circular Couette flow undergo inertial migration to an equilibrium position near the middle of the annular gap.
Majji and Morris ~\cite{majji2018inertial}, using the same TC apparatus as Majji, Banerjee, and Morris ~\cite{majji2018inertial_JFM}, studied inertial migration of particles under dilute conditions in three flow structures: CCF, TVF, and WTV. The particle volume fraction was kept at $\phi=0.001$ so that flow structures and flow transition Reynolds numbers were those of the Newtonian fluid. Particles migrate to different locations depending on the flow structure, as shown in Fig.~\ref{fig:migration-dilute}. 

In CCF ($Re = 83$), initially uniformly distributed particles in the annular region migrated to a location near the mid-radial plane, but slightly closer to the inner cylinder (0.4$\delta$ from the inner cylinder). Inertial migration was a slow process, as particle migration velocities were two orders of magnitude smaller than the linear velocity of the inner cylinder ($100V_m\approx r_i \omega_i$) and particles took up to $\approx 1407$ inner cylinder rotations, or $\approx 766t_d$ where the diffusion time is $t_d = {\rho\delta^2}/{\mu}$, to reach the steady-state location. The azimuthal velocity profile of CCF has two terms varying as $r$ and $1/r$. The first term with uniform shear rate causes the particles to migrate to the mid-radial plane, whereas the second term with non-zero shear gradient offsets the equilibrium position slightly towards the inner cylinder. Increasing the radius ratio of the TC setup should increase this offset.

An increase in particle concentration to $\phi=0.01$ in CCF led to formation of a band of particles instead of a line near the equilibrium location in the $r$-$z$ plane~\cite{baroudi2020effect}. In their 2019 study, Ramesh, Bharadwaj and Alam~\cite{ramesh2019suspension} measured flow velocity profiles using PIV up to $\phi = 0.15$, beyond which concentration the index matching was too poor for accurate measurements.  The PIV results were remarkable for showing that particle-laden flows showed a shear-banding type velocity profile in CCF, which the authors ascribed to particle migration toward a position near the annular center, as shown by Majji \& Morris~\cite{majji2018inertial}. This migration requires an extended period under similar kinematics and is associated with hysteresis; this point was explored and the validity of the Ramesh {\em et al.} assertion of migration influence was placed on sound footing by use of different flow protocols in which migration was controlled by Baroudi {\em et al.} ~\cite{baroudi2020effect}. 

In TVF ($Re = 144$), Majji and Morris ~\cite{majji2018inertial} observed that initially uniformly distributed particles in each of the Taylor vortices were convected by the secondary flow and simultaneously pushed away from the wall and vortex center, leading to regularly spaced accumulations within a circular region in the $r$-$z$ plane. The exact shape of the accumulation was affected by the strength of the vortex.  At lower $Re$, particles form a fairly uniformly distributed circular disk near the vortex center instead of a circular ring due to weaker gradients inside the vortex. At higher $Re$, regularly spaced counter-rotating pairs of particle rings were observed, but the rings within each pair were much closer. Their measurements in WTV ($Re > 151$) showed the particles remained well mixed as the azimuthal waviness of the Taylor vortices has a much shorter time scale compared to the particle inertial migration time scale, preventing the particles from having sufficient time to migrate. For the same reason, particles stay fairly uniformly distributed at low and moderate concentrations in all of the non-axisymmetric flow structures: WSV, SVF, and RIB. 

Ramesh, Bharadwaj and Alam~\cite{ramesh2019suspension} further explored the vortex structure in TVF with concentrations between 0.05$\le\phi\le$0.15 using PIV measurements. The particles were observed to significantly alter the TVF structure: the vortex centers were offset from the mid-radial plane in an alternating fashion along the axial direction, a marked set of differences from Newtonian fluids. 

This radial offset of vortices in TVF, for $\phi = 0.05$ - 0.15, was noted by Ramesh {\em et al.}, but the fact this is a different flow state (work in preparation  M. Majji/J. Morris, unpublished) was not mentioned.  The new vortex structures for suspension resemble the classic Newtonian TVF when viewed in the z-$\theta$ plane near the outer cylinder. The axial inter-vortex transport resembles the classic Newtonian spiral vortices but with a key difference: in Newtonian spirals, all the spirals in the axial direction are connected to have axial transport with vortex centers aligned; whereas in the new structures generated by the suspension, each counter-rotating pair is axially connected for transport with the two vortex centers offsetting from the mid radial plane.  

Overall, the inertial migration of particles in axisymmetric flow structures, such as CCF and TVF, leads to accumulation of particles in equilibrium locations at dilute concentrations and non-uniform concentration profiles in the annular region at higher concentrations, while the particle concentration apparently remains much more uniform across the flow structure in NAF structures such as WTV, SVF, RIB, and WSV. 

\begin{figure}
  \centering
    \includegraphics[scale=1.0]{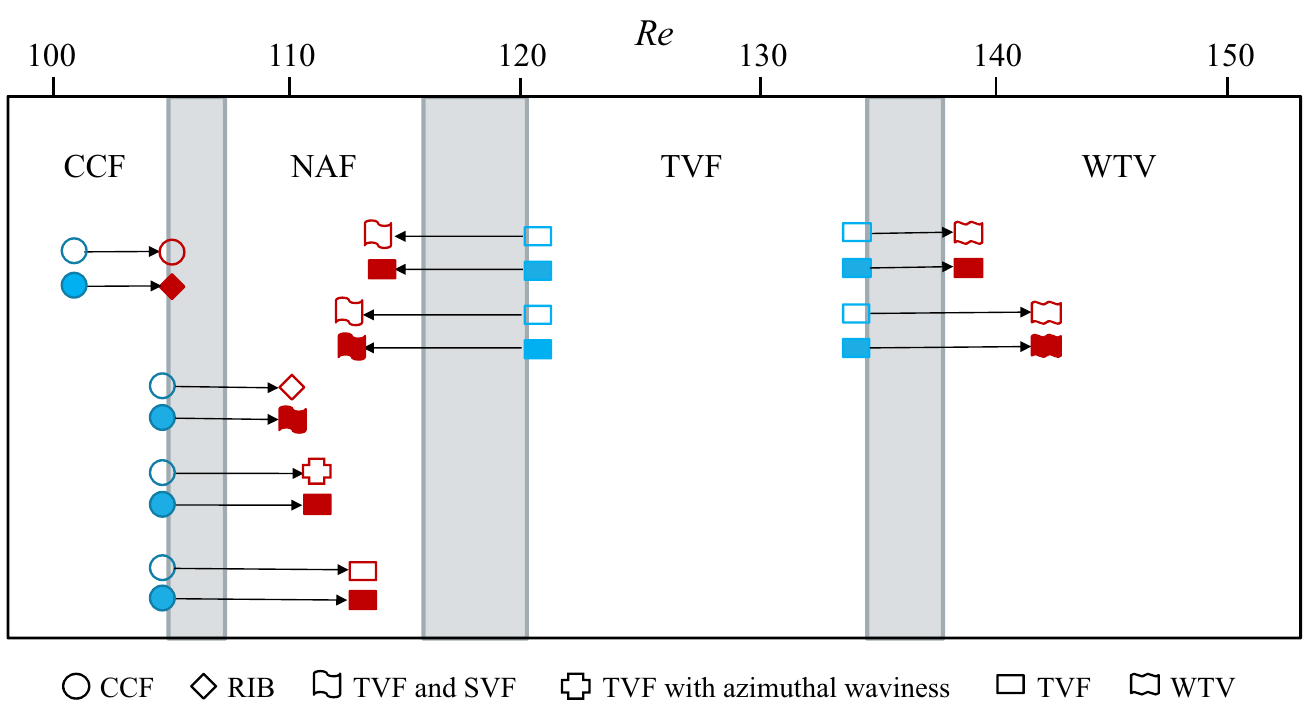}
  \caption{Flow transition map for rapid $Re$ step change experiments showing the Reynolds number corresponding to various flow structures and transitions for $\phi = 0.10$ suspension with $\alpha = 30$. Gray shaded regions show the range of $Re$ over which flow transition away from the initial flow state is observed. The initial and final states for different experiments are marked on the flow transition map. Open symbols correspond to experiments with uniform suspension concentration at the initial state and filled symbols to experiments with inertially migrated particle concentration at the initial state. Abbreviations: CCF, circular Couette flow; NAF, non-axisymmetric flow, TVF, Taylor vortex flow; WTV, wavy Taylor vortices; RIB, Ribbons; SVF, spiral vortex flow. Adapted from Baroudi {\em et al.}~\cite{baroudi2020effect}.}\label{fig:map_Re_step}
\end{figure}

The particle concentration being uniform or non-uniform in the flow structure just before the transition depends on whether it is a decreasing $Re$ or an increasing $Re$ procedure and thus can lead to significant hysteresis. Using the same TC setup as Majji {\em et al.}~\cite{majji2018inertial_JFM}, Baroudi, Majji and Morris~\cite{baroudi2020effect} studied the influence of inertially-migrated non-uniform concentration profiles on flow transitions. 
Considering $\phi=0.1$, and $\alpha = 30$, quasi-steady changes in $Re$ resulted in the flow transition sequence WTV-TVF-SVF-RIB-CCF (Majji {et al.}~\cite{majji2018inertial_JFM}). Baroudi {\em et al.} focused on the CCF-NAF, TVF-NAF and TVF-WTV transitions and measured the change in transition $Re$ and the type of flow states post transition for (I) uniform and (II) inertially-migrated non-uniform concentration profiles in the flow structures just before transition. To achieve this they started with a flow structure close to the transition boundary (initial state at $Re_1$) and quickly changed the $Re$ to cross the transition boundary (final state at $Re_2$). Uniform and inertially-migrated concentration profiles were established by quickly increasing the $Re$ from a quiescent well-mixed state to the desired initial state and waiting there for short time ($\approx 1$ min) and sufficiently long time ($\approx 50$ min), respectively, before changing $Re$. Fig.~\ref{fig:map_Re_step}, adapted from Baroudi {\em et al.}, shows the flow transition map for rapid $Re$ step-change experiments.   

The authors~\cite{baroudi2020effect} concluded that the inertially-migrated concentration profile destabilizes the CCF structure at lower $Re$ compared to the uniform concentration profile case and affects the observed flow states post transition based on the following experimental observations. When the $Re$ was increased from $Re_1=100$ to $Re_2=105$, the flow remained in the initial CCF state for the uniform concentration profile case, but transitioned into a RIB state for the non-uniform concentration profile case. For the $Re_1 =100 \to Re_2=110.1$ case, CCF transitioned into RIB and SVF for the uniform and non-uniform concentration profile cases respectively. For the $Re_1 =104.6 \to  Re_2 =111$ case, CCF$\to$SVF and CCF$\to$TVF transitions were observed for uniform and non-uniform concentration profiles and finally for the $Re_1=104.6 \to Re_2 =113.1$ case, CCF$\to$TVF transition was observed for both the concentration distributions.

In contrast, the inertially-migrated concentration profile stabilizes the TVF structure for both TVF$\to$NAF and TVF$\to$WTV transitions compared to the uniform concentration case. For the TVF$\to$NAF transition case, when the $Re$ was decreased from $Re_1=120.5 \to Re_2 =113.7$, TVF transitioned into SVF for the uniform concentration profile case and stayed the same for the non-uniform concentration case. For the $Re_1=104.6 \to Re_2=112.5$ case, TVF$\to$SVF transition was observed for both the concentration profiles. 

For the TVF$\to$WTV transition case ($Re_1 =131.4 \to Re_2=138.7$), the TVF structure transitioned into the WTV structure for the uniform concentration case and stayed the same for the non-uniform concentration profile case. For the $Re_1 =134.1 \to Re_2=142.1$ case, TVF$\to$WTV transition was observed for both $\phi$ profiles.

Based on these experimental observations, we may attribute at least part of the hysteresis behavior to migration.  In particular, hysteresis observed in the transition $Re$ between the quasi-steady decreasing and increasing $Re$ protocols used in Majji, Banerjee and Morris~\cite{majji2018inertial_JFM} and Ramesh, Bharadwaj and Alam~\cite{ramesh2019suspension} to the inertially-migrated $\phi$ in axisymmetric flow structures and more uniform $\phi$ in the NAF structures. Note that the transition $Re$ for all the flow transitions, with uniform and non-uniform concentration profiles, studied in this work were lower than those of a Newtonian case.

\subsection{Inner cylinder torque measurements} \label{torqueMeasurements}

Torque measurements of suspension TCF were first explored by Ramesh {\em et al.}~\cite{ramesh2019suspension}. Torque measured on the inner cylinder was used to calculate a pseudo-Nusselt number, $Nu_{\omega}$, which relates the dimensionless torque ($G$) at a given Reynolds number to a calculated laminar value, $G_{lam}$:  
 \begin{equation}
 Nu_\omega = \frac{G}{G_{lam}}, \label{eq:Nu}
 \end{equation}
where
 \begin{equation}
 G = \frac{\tau \rho}{2\pi L \mu(\phi)^2},
 \end{equation}
 and
  \begin{equation}
 G_{lam} = \frac{\eta}{(1-\eta)^2}Re(\phi),
 \end{equation}
where $\tau$ is the measured torque, $\rho$ is the density, $L$ is the inner cylinder height, $\mu(\phi)$ is the effective viscosity of the suspension, and $\eta$ is the radius ratio of the TC device. The laminar dimensionless torque value assumes CCF with uniform particle distribution in the annular region for infinite cylinders, and is only a function of $Re$ and $\eta$. In this form, $Nu_\omega$ represents a dimensionless angular momentum transfer parameter that measures the effectiveness of the transverse convective angular velocity transport in terms of purely molecular transverse transport (Eckhardt {\em et al.}~\cite{eckhardt2007torque}).
Ramesh {\em et al.}~\cite{ramesh2019suspension} showed that the primary bifurcation away from CCF results in a nearly discontinuous slope increase in the $Nu_\omega$-$Re$ curve, behavior that is well-known in the rheological community. A mild change in the slope was observed in suspensions with $\phi = 0.1$ and $\phi = 0.2$ at the transition boundary from NAF states to TVF. For the TVF-WTV transition, the slope change was not distinct. 

Additional torque measurements were conducted by Moazzen {\em et al.}~\cite{moazzen2022torque} in a rheometer-based device.  This study focused on characterizing the torque behavior in different flow states and analyzing the unsteady features of NAF structures. Flow transitions were identified by image analysis and torque measurements; the authors considered the torque $\tau$ in dimensionless form as the Nusselt number (Eq.~\Ref{eq:Nu}). For low particle loading, $\phi < 0.06$, the sequence of flow transitions resembles that of a pure fluid, namely CCF $\leftrightarrow$ TVF $\leftrightarrow$ WTV. The $Nu_\omega$-$Re$ curve showed a sharp change in the slope at the onset of the TVF transition and a less distinct change for the TVF-WTV transition. At $\phi \geq 0.06$, both ramp up and ramp down tests resulted in the flow transition sequence CCF $\leftrightarrow$ SVF $\leftrightarrow$ TVF $\leftrightarrow$ WTV. Fluctuations and smooth changes in the slope of the $Nu_\omega$–$Re$ curve were observed as the flow transitioned from CCF to TVF via SVF. The non-axisymmetric structure, SVF, observed in their study is similar to those observed by Majji {\em et al.} and Ramesh {\em et al.}; however, unlike in Ramesh {\em et al.} with very similar conditions, the authors did not observe co-existing states of stationary (TVF) and traveling (WTV or SVF) waves. Increasing particle concentration led to a general decrease in the critical $Re$ of the primary bifurcation, in agreement with previous experimental observations. The critical $Re$ for the transition to WTV was constant for $\phi<0.15$ and sharply decreased for higher concentrations. A hysteretic behavior was observed for primary bifurcations of suspensions with $\phi>0.12$ and the secondary and tertiary bifurcations for $\phi>0.03$. The observed hysteretic behavior is in accord with previous experimental studies~\cite{majji2018inertial_JFM,ramesh2019suspension,baroudi2020effect}. The range of $Re$ corresponding to SVF increased until $\phi = 0.15$ and then reduced in both directions of ramping; previous studies~\cite{majji2018inertial_JFM,ramesh2019suspension} did not report such non-monotonic behavior of the $Re$ range over which the SVF state exists. The $Re$ span of the TVF decreased with increasing concentration in good qualitative agreement with  Majji \textit{et al.} for $\phi\leq 0.2$.

\begin{figure}
  \centering
    \includegraphics[scale=0.95]{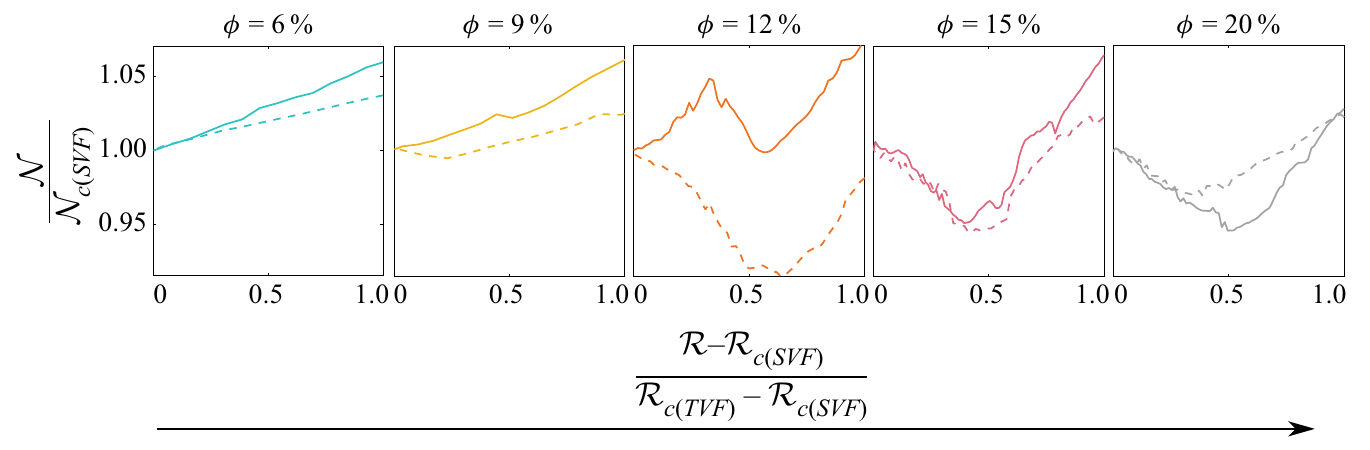}
  \caption{Variation of the Nusselt number, denoted as $\mathcal{N}$ in this figure, in the spiral vortex flow state (SVF). The Nusselt number is normalized by critical Nusselt number of SVF state for each concentration as a function of dimensionless Reynolds number for $\phi = 0.06,0.09,0.12,0.15,0.20$. Here, $\mathcal{R}_{c(SVF)}$ is the critical Reynolds number of the SVF state and $\mathcal{R}_{c(TVF)}$ is the critical Reynolds number of Taylor vortex flow (TVF) state. Solid and dashed lines represent the ramp-up and ramp-down protocol, respectively. Adapted from Moazzen {\em et al.}~\cite{moazzen2022torque}.} \label{fig:torque}
\end{figure}

An overall enhancement in the torque exerted on the inner cylinder was observed with increasing $\phi$. However, the evolution of the Nusselt number with $Re$ (or equivalently Taylor number: $Ta = (1+\eta)^6Re^2/(64\eta^4) $) exhibited dependence on the flow structures. As opposed to the general increasing trend of $Nu_\omega$ with $Re$ in TVF and WTV regimes, the changes of $Nu_\omega$ with $Re$ corresponding to SVF for $\phi>0.06$ were non-monotonic and globally constant or decreasing at higher $\phi$, as shown in Fig.~\ref{fig:torque} reprinted from Moazzen {\em et al.~\cite{moazzen2022torque}}. The authors ascribed this behavior to the interplay between the decay of the radial convective momentum transport caused by the axially traveling vortices and the increase in the kinetic energy with increasing rotation rates. Detailed analysis of the Nusselt number scaling was reported as a function of Taylor number, as $Nu_\omega \propto Ta^{\alpha}$, for different particle concentrations in TVF and WTV. In TVF, as $\phi$ increased to $0.15$, the scaling exponent decreased from $\alpha \simeq 0.6$ for pure fluid to $0.49$. For $\phi>0.15$, the trend is reversed with $\alpha = 0.53$ for $\phi = 0.2$. For WTV, when increasing $\phi$ up to $0.12$, the scaling exponent decreased from $0.3$ in the case of pure fluid to $0.18$. The trend is then reversed with an increase in the scaling exponent for $\phi>0.12$. It is worth noting that for higher values of $Ta$ in the WTV regime, Dash {\em et al.}~\cite{dash2020particle} observed the following scaling $Nu_\omega\propto Ta^{0.23}$, where the scaling exponent is independent of the particle loading. On the other hand, the scaling exponent of $Ta$ in the experiments of Ramesh {\em et al.}~\cite{ramesh2019suspension} (as seen in their figure 27), while not reported explicitly, shows a dependence on particle loading. 

Based on the observed scaling exponents, Moazzen {\em et al.}~\cite{moazzen2022torque} concluded that in TVF and WTV states, the presence of particles reduces the transverse momentum transfer for $\phi<0.15$ and $\phi<0.12$, respectively, and this behavior is reversed for higher $\phi$. This interpretation does not agree with the enhanced radial transport in TVF with increasing particle loading for $\phi\le0.15$ reported by Ramesh {\em et al.} based on their PIV measurements. It is difficult to come up with a mechanistic understanding of the momentum transfer behaviour in suspension based on the reported scaling for $Nu_\omega$. The Nusslet number is defined in terms of a calculated “laminar” value of the dimensionless torque ($G_{lam}$) assuming CCF regime for homogeneous suspension in TCF with infinite cylinders. This might not hold due to particle migration in CCF and the development of inhomogeneous distribution of particles across the gap which lead to changes in the local shear rate near the inner cylinder. Discrete particle numerical simulations will help obtain better insights into the relative importance of different mechanisms of radial momentum transfer in suspensions TCF.

Temporal-spectral analyses were performed by Moazzen {\em et al.}~\cite{moazzen2022torque} to characterize the unsteady features of the NAF states. Their analysis revealed that the characteristic frequency of the SVF was always half that of the inner cylinder $f_{SVF}/f_{\omega_i}\approx 0.5$ for all concentrations and did not change with $Re$. However, for WTV, a nontrivial frequency behavior was reported: at the onset of wavy vortices, the relative azimuthal wave frequency ($f_{WTV}/f_{\omega_i}$) was observed to decrease with concentration until $\phi = 0.12$, and the trend is reversed for $\phi > 0.12$. Furthermore, the relative frequency $f_{WTV}/f_{\omega_i}$ decreased with increasing $Re$ for all concentrations except $\phi = 0.12 $ and $0.15$. This trend differs from the observation of Ramesh \textit{et al.} (2019), where the characteristic frequency associated with WTV was approximately half that of the inner cylinder, with negligible dependence on concentration for $\phi\leq 0.2$. 

\subsection{Lower-order flow transitions in the counter-rotation regime} \label{lowCounterRegime}

In pure Newtonian fluids, experiments have shown that counter-rotation increases the number of available flow states; the flow state map (Fig.~\ref{fig:setup-map}) reproduced from Andereck \textit{et al.} ~\cite{andereck1986flow} shows a large number of new states as $Re_{o}$ changes and it is expected that suspension flows will lead to an equally or even more diverse state map. 

Singh {\em et al.}~\cite{singh2022counter} conducted a study of flow transitions with counter-rotating cylinders using a rheometer-based device.  The inner and outer cylinder were independently controlled, and the level of counter-rotation was characterized using the speed ratio $\Omega = \omega_o / \omega_i$.  Testing was conducted by varying the rotation rate of the inner and outer cylinders while maintaining a constant $\Omega$.  Five levels of counter-rotation were tested:  $\Omega$ = 0 (only inner cylinder rotation), -0.1, -0.25, -0.5, and -1.0.

The authors validated their apparatus by comparing pure fluid up-ramp and down-ramp transition $Re$ with those of Andereck \textit{et al.} ~\cite{andereck1986flow}.  Despite differences in aspect ratio ($\Gamma = 8.46$ versus $\Gamma = 30$) and testing protocol (holding $\Omega$ constant versus holding the $\omega_o$ constant), the evolution of flow states and transition $Re$ were largely similar between the two experimental setups.  In general, the smaller aspect ratio resulted in higher transition $Re$, a result previously observed in Ramesh, Bharadwaj and Alam~\cite{ramesh2019suspension}.  
A wide parameter space was tested and the authors extensively characterized changes in flow state, frequency response, number of vortex rolls, and torque.  While it is difficult to generalize all of the results, the authors found that for all $\Omega$, increasing $\phi$ decreased the $Re$ of the primary transition, in agreement with the destabilizing effect observed in studies with only inner cylinder rotation.  This destabilizing behavior was also observed for most of the secondary and tertiary transitions, with some exceptions across the tested parameter space. The level of counter-rotation was found to affect the evolution of flow states.  For all tests with lower levels of counter-rotation ($\Omega \geq -0.5$), the primary transition occurred between CCF and TVF, followed by wavy flow states.

Tests with the highest level of counter-rotation ($\Omega = -1.0$) showed a transition directly from CCF to an oscillatory spiral vortex flow, a state with a spiral wave propagating downward at approximately the frequency of the inner cylinder rotation.  The flow further transitioned to ISV, followed by a “non-propagating interpenetrating state.”  The authors note that a similar flow state evolution has been previously observed in pure fluid TC flows (Andereck \textit{et al.} ~\cite{andereck1986flow}). 

Torque measured on the inner cylinder was used to calculate $Nu_{\omega}$.  All operating conditions showed a general increase in $Nu_{\omega}$ with increasing $Re$; however, specific changes in the flow state were accompanied by discontinuous changes in $Nu_{\omega}$. For example, increasing the number of vortex rolls in wavy Taylor flow led to a sharp increase in $Nu_{\omega}$. For $\Omega = -0.5$, transitions between wavy states, with a singular dominant frequency, to modulated wavy states were accompanied by discontinuous drops in $Nu_{\omega}$, indicating a decrease in torque that the authors attribute to modulation waves “costing” energy for the flow.

Torque measurements for $\Omega = -1.0$ in ISV showed significant fluctuations in $Nu_{\omega}$ that persisted to the highest $Re$ tested. These fluctuations were attributed to the asymmetric interaction of the spirals producing fluctuations in the torque on the inner cylinder.  In addition, the authors observed that the level of fluctuation tended to decrease with increasing $\phi$, indicating that the particles damp the spirals as they interpenetrate each other through the flow.

\subsection{Higher-order flow transitions} \label{highOrderRegime}

The work of Dash,  Anantharaman \& Poelma~\cite{dash2020particle} considered higher $Re$ and $\phi$ than the prior studies by Majji {et al.} and Ramesh {\em et al.}: $Re \approx 2000$ for the pure liquid or low $\phi$ and $\phi \le 0.4$, $Re/Re_c \approx 2.5$ for $\phi = 0.4$, where $Re_c$ is the critical Reynolds number of the primary transition.  This extended the regime maps for suspension to include modulated WTV and wavy turbulent vortex flow. In addition, there was a scaling of torque and use of the power spectrum of the reflected light from small regions of the flow.  This study was not as closely temperature controlled as the previous studies, so with the temperature influence on viscosity and loss of neutral buoyancy as well as higher ramp rates to allow consideration of the wider range of $Re$, some precision on phase boundaries was necessarily sacrificed.  However, the qualitative behavior in terms of the influence of solid fraction of these phase boundaries, including the nonaxisymmetric states at the primary bifurcation from CCF of Majji {\em et al.}~\cite{majji2018inertial_JFM} and the co-existing states of Ramesh {\em et al.}~\cite{ramesh2019suspension}, is consistent with earlier work.

In terms of new states, the authors identified an azimuthally localized (i.e. for only a fraction of the angle $\theta$ about the cylinder axis) waviness in the boundary between TVF and WTV.  The waviness occurs on what is otherwise apparently a Taylor vortex, and was found to persist for hours, up to about 15,000 inner cylinder rotations, effectively ruling out the possibility of this as a transient.

The spectral analysis showed significant complexity in the range of $Re/Re_c(\phi) = 1$ - 1.5, while higher $Re$ and higher-order transition showed a simpler behavior, albeit one with significant hysteresis.  The complexity in this regime for $\phi = 0.15$ and 0.3 especially shows this, in their Figs 12 and 13, respectively.  This suggests that the influence of neutrally-buoyant particles on the detailed form of the dynamical system, at least for these moderate volume fractions, may be qualitatively strongest in the primary and secondary flow transition region, where the destabilization of the CCF by particle loading and triggering of nonaxisymmetric states as primary bifurcations in the inner cylinder-driven flow (whereas it is seen only with counter-rotating cylinders for pure fluid) was established by earlier work described above.  The spectral analysis of Dash {et al.} thus provides a more encompassing global view of the role of particles on the response of the inner-cylinder-driven system. 

The authors considered the torque $T$ in dimensionless form as the Nusselt number $Nu_{\omega}$. The scaling for the behavior beyond the transition to wavy vortices, i.e. in the higher-order transition regime, was shown to satisfy $Nu \sim Ta^{0.24} \eta_r^{0.41}(\phi)$ 
where $\mu_r = \mu_s(\phi)/\mu$ is the viscosity of the suspension relative to the pure suspending fluid.

In a forthcoming study, Baroudi and Peluso (to be published) experimentally tested the effect of particle concentration on turbulent transitions in water-glycerol mixtures with polystyrene particles.  The apparatus used was similar to the design of Majji {\em et al.}~\cite{majji2018inertial_JFM}, except with an increased cylinder height, leading to an aspect ratio of $\Gamma = 38$.  Volume fractions up to $\phi$ = 30\% were tested following a ramp-down protocol and images and videos of the flow, combined with temporal- and spatial-spectral analysis, were used to classify the flow state for each Reynolds number. The device and testing protocol were validated by comparing the general behavior of the flow in the laminar regime with previous studies. The primary objectives of the study were to characterize the flow states and their progression for suspensions in the turbulent regime and to determine the effect of concentration on higher-order turbulent transitions.  Turbulent Taylor vortex flow (TTV) was achieved for all concentrations, with higher concentrations requiring higher $Re$ to reach TTV.  For example, pure fluid suspensions transitioned out of the TTV state at approximately Re = 2200, while the 30\% particle suspension transitioned at approximately Re = 3600.  During ramp down tests, all suspensions transitioned to a wavy turbulent state, followed by a series of sub-states within the wavy turbulent regime.  These sub-states were denoted by the flow exhibiting different number of vortex pairs and azimuthal waves.  Suspension concentration affected the $Re$ ranges for each of these sub-states; however, the frequency of the main wave, normalized by the inner cylinder rotation frequency, followed the same progression for all concentrations.    

As $Re$ decreased further, the flow transitioned from a wavy turbulent state to a laminar wavy state.  For pure fluid and low-concentration suspensions, the background noise in the temporal spectra decreased as the flow entered the laminar regime and this was accompanied by a transition in the normalized frequency of the wave.  For higher concentration suspensions, a shift in the background noise could not be detected, as the increased number of particles obscured the finer scale fluctuations.  However, the same abrupt transition in relative frequency was observed for all suspensions, allowing this transition to be used to characterize the shift between turbulent and laminar flow.  Opposite to the transitions out of TTV, higher concentration suspensions transitioned to laminar flow at lower $Re$, indicating that particle addition destabilizes the flow transitioning between the turbulent and laminar states.

\section{Modelling and linear stability analysis}\label{LSA-Modelling}

Ali \textit{et al.}~\cite{ali2002hydrodynamic} carried out linear stability analysis (LSA) to probe the stability of the CCF of a non-Brownian hard sphere suspensions in the dilute limit $\phi \leq 0.05$. The conservation equations for the system were based on a two-fluid formulation under the assumptions of a very dilute suspension $\phi \ll 1$ and a small particle Reynolds number $Re_p<1$. The stability of CCF with respect to axisymmetric perturbations that are periodic in the axial direction with a wave number $k_z$ was examined for different particle sizes with a gap-to-particle size ratio of $50\leq\alpha \leq 5000$, and different particle-to-fluid density ratios $\epsilon = \rho_p/\rho_f $ ($\epsilon= 0.001, 1, 10, 833$).  Their analysis predicted that, at a given radius ratio, the critical Reynolds number ($Re_c$) of the primary instability decreases as the particle concentration increases up to $\phi = 0.05$, and $Re_c$ was independent of the particle size. Moreover, they observed that increasing the ratio of particle-to-fluid density above 1 increases the degree of destabilization of the CCF. The critical axial wave number was the same for a suspension as for a pure fluid ($\delta k_{z}/\pi \approx 1$), and was independent of particle density and size ratios. In the same study, the authors conducted experiments for suspension of neutrally buoyant particles for concentrations $\phi \leq 0.005$ and particle size ratio of $\alpha \approx 260$ in a TC cell with an aspect ratio of $\Gamma = 96$ and a radius ratio $\eta = 0.824$. Their experiments showed a stabilizing effect of the particles, in contrast to the predictions of their LSA. 

For higher particle concentrations $0.0\leq \phi \leq 0.5$, Gillissen and Wilson~\cite{gillissen2019taylor} investigated the linear stability of the circular Taylor-Couette flow (CCF) of a suspension of non-Brownian neutrally buoyant spheres w.r.t. axisymmetric perturbations. In their LSA,  the suspension was treated as an effective fluid, governed by momentum balance and continuity equations, with rheological properties depending on particle concentration. In this approach, the solid particles and the fluid move together as a single phase with identical averaged velocity. To describe the rheology of the suspension, the authors used their previously developed particle stress constitutive model for shear rate-invariant dense suspensions ($0.2 \leq \phi \leq 0.5$) of non-Brownian spheres at vanishing  $Re_p$ (Gillissen \& Wilson~\cite{PhysRevE.98.033119}). In this model, the extra stress is induced by the lubrication forces between the particles, and all non-hydrodynamic forces are ignored. The non-Newtonian feature of the suspension due to the shear-induced anisotropic suspension microstructure is characterized by the second normal stress difference, while the first normal stress difference is assumed to be zero. 

For a flow driven by inner cylinder rotation with a radius ratio of $r_i/r_o = 0.877$, similar to Majji \textit{et al.}~\cite{majji2018inertial_JFM}, Gillissen and Wilson~\cite{gillissen2019taylor} examined the stability of CCF to infinitesimal disturbances that are axisymmetric and periodic in the axial direction with a wave number $k_z$. They predicted the critical Taylor number ($Ta$) defined based on the effective suspension viscosity as a function of $\phi$. They found that the most unstable axisymmetric mode was nonoscillatory with a critical axial wave number of $\delta k_{z}/\pi \approx 1$. Their analysis showed a negligible effect of the particles on the stability of CCF for $\phi\leq 0.2$ and a destabilizing effect for $\phi\geq0.2$. 
For $\phi\leq 0.2$, the effective fluid approach with the particle stress constitutive model~\cite{PhysRevE.98.033119} used in their analysis did not agree with the experimentally observed destabilizing effect of the particles in this concentration range (Majji \textit{et al.}~\cite{majji2018inertial_JFM},  Ramesh \textit{et al.}~\cite{ramesh2019suspension}, \& Moazzen {\em et al.}~\cite{moazzen2022torque}). For $\phi \geq 0.2$, their results showed less destabilization compared to the experiments of Majji \textit{et al.}, as seen in Figure 1 of their work. 
The discrepancy between the experimental data and the result of their study was attributed to the neglect of sphere inertia in their model and the axisymmetry of the instability modes utilized in their analysis as opposed to the non-axisymmetric form of the primary instability, e.g., ribbons and spirals, observed in experiments for $\phi\geq 0.05$. Additionally, the effective fluid approach adopted in their work can not capture the relative motion between the fluid and the particle phases due to particle migration which has been shown to affect flow transitions and observed flow structures~\cite{baroudi2020effect}.

Kang and Mirbod~\cite{kang2021flow} conducted numerical simulations using the continuum two-phase “suspension balance” approach, consisting of mass and momentum balances for the bulk suspension and particle phase. The balance equations were discretized using the finite volume method in a computational domain with a radius ratio of  $\eta = r_i/r_o = 0.877$. 
Periodic boundary conditions were applied in the axial direction, and a no-slip boundary condition was imposed on the cylindrical surfaces. Two particle sizes with gap-to-particle diameter ratios $\alpha = \delta/ {d_p} = 30$ and $100$ were considered, similar to the experiments of Majji, Banerjee \& Morris~\cite{majji2018inertial_JFM}, and the solid volume fraction was fixed at $\phi = 0.1$. The flow and particle concentration fields were examined in different flow regimes. In their simulations, Kang and Mirbod~\cite{kang2021flow} assumed that the inertia of particles is negligible $Re_p\ll1$, and the shear-induced migration driven by the viscously generated particle normal stresses is dominant, while the inertial migration of the particles is neglected. It is worth noting that the assumption of negligible inertial migration conflicts with the observations reported in previous experimental studies (Majji \& Morris~\cite{majji2018inertial_JFM}; Ramesh \textit{et al.}~\cite{ramesh2019suspension}; Baroudi, Majii \& Morris~\cite{baroudi2020effect}).   

In the CCF regime, the predicted azimuthal velocity profile $u_{\theta}(r)$ was linear and similar to that of the pure fluid. Moreover, the particle concentration profile $\phi(r)$ was also linear, and particles were found to migrate radially outward for both particle sizes. The predicted azimuthal velocity profile observed in the CCF regime differs from that reported by Ramesh \textit{et al.}\cite{ramesh2019suspension} based on their PIV measurements. As mentioned earlier, Ramesh \textit{et al.} showed that the particles produced a nonlinear (shear-banded-type) profile in CCF and these changes to the velocity profile were ascribed to particle migration, as demonstrated by Majji \& Morris~\cite{majji2018inertial}. 

As $Re$ increased, Kang and Mirbod found that the primary transition away from CCF occurred at the same effective $Re$ for both particle sizes, and the particles had no impact on the primary instability's critical conditions compared to pure Newtonian fluid. This behavior conflicts with the particle-induced destabilization seen in the experimental studies discussed in the previous section~\cite{majji2018inertial_JFM,ramesh2020interpenetrating,ramesh2019suspension,moazzen2022torque}. The discrepancies between their results and the reported experimental findings were ascribed to the neglect of particle inertia and inertial migration in their simulations.

The flow transitions for the suspension of the smaller particles with $\alpha = 100$ followed CCF $\rightarrow$ TVF $\rightarrow$ WTV with increasing Re, as in the case of pure fluid. 
While particles were almost uniformly distributed in WTV, shear-induced particle migration to the center of the vortices was observed in TVF.
For the suspension with $\alpha =30$, a non-axisymmetric flow state (SVF) was found to be a primary bifurcation from CCF, similar to the experimental observations~\cite{majji2018inertial_JFM,ramesh2019suspension,moazzen2022torque}. However, the observed flow transitions sequence of CCF $\rightarrow$ SVF $\rightarrow$ WSV $\rightarrow$ WTV did not match that reported in the experiments at $\phi = 0.1$. The axisymmetric TVF state did not develop after the non-axisymmetric state, and the transition to the WTV occurred at a higher $Re$ than pure fluid. Higher particle concentration was realized in the core of the vortices with positive azimuthal vorticity in the SVF state as a result of shear-induced migration. Due to the WSV's oscillation, particles were dispersed more broadly in the region of wavy vortices, and less accumulation was seen in the vortex core.

The authors compared the torque exerted on the inner cylinder by the suspension and pure fluid using the effective-Nusselt number, following a similar approach as Ramesh \textit{et al.}~\cite{ramesh2019suspension}. In agreement with previous experimental studies, the dimensionless torque ($G$) was found to increase with $Re$ with an enhancement of the torque with particle loading and the slope of $G-Re$ and $Nu_{\omega}$-$Re$ curves changed at the onset of different flow transitions. Following the primary transition from CCF, $Nu_{\omega}$ increased rapidly with $Re$, yet, the values of suspensions were observed to be a bit lower than those of a pure fluid flow. In addition, the values of $Nu_{\omega}$ for the suspension of larger particles with $\alpha = 30$ at the SVF state were slightly smaller than those of $\alpha = 100$ at TVF state at the same $Re$.  The effective Nusselt numbers of pure fluid and suspensions were the same in the WTV regime. The authors concluded that the axial traveling wave of spiral vortices weakens the convective momentum transfer in the radial direction and reduces the effective Nusselt number and torque acting on the inner cylinder.

\section{Perspective}\label{Perspective}

The study of particle-laden TCF has drawn attention in recent years. This flow undergoes several transitions before reaching turbulence when the inner cylinder's angular velocity exceeds that of the outer cylinder, making it an ideal setting for examining the stability and fluid mechanical behaviour of inertial suspensions. The work discussed in this review demonstrates that particles alter inertial flow transitions and structures known for pure fluids in Taylor-Couette flow and establishes the influence of the two-phase nature of suspension due to particle migration and the ensuing history effects on the observed behavior. However, a quantitative explanation of the origin of the observed behavior is still lacking. The reported deviations of the suspension behavior from a pure fluid cannot be explained by scaling the Reynolds number with the effective suspension viscosity to account for the additional dissipation caused by the particles. Furthermore, rheological properties like normal stress differences that do not arise from inertia mechanisms are probably insignificant at low concentrations and cannot account for the behavior seen for $\phi<0.2$; however, they might be influential at higher concentrations. Experimental evidence of suspension behavior approaching that of a pure fluid with particle size reduction indicates that the particle-induced inertial stresses may be a potential candidate for explaining the suspension behavior. Recent attempts to analyze and model particle-laden TCF~\cite{gillissen2019taylor,kang2021flow}, while capturing some qualitative features, were partly unsuccessful in reproducing or explaining the experimental findings as the available continuum descriptions are not yet adequate to describe the behavior of inertial suspensions. Thus, there is a clear need for developing continuum models and improving existing ones to enable reliable predictions of the flow behavior of inertial suspensions.
Additionally, fully resolved multiparticle numerical simulations~\cite{maxey2017simulation,de2019assessment} could be utilized to investigate the mechanism underlying the destabilizing effect of the particles. Such simulations can also provide a fundamental understanding of the processes by which mass and momentum are transported in various flow structures in suspension, which will help in the development of predictive models to describe inertial suspension flows. Model predictions can be validated against the experimental observations in TC geometry, and prediction quality can be systematically analyzed to drive model improvement and focus of experiments.  

\bibliographystyle{ieeetr}
\bibliography{Review_Refs.bib}

\end{document}